\documentclass{article}

\usepackage{PRIMEarxiv}

\usepackage[utf8]{inputenc} 
\usepackage[T1]{fontenc}    
\usepackage{hyperref}       
\usepackage{url}            
\usepackage{booktabs}       
\usepackage{amsfonts}       
\usepackage{nicefrac}       
\usepackage{microtype}      
\usepackage{lipsum}
\usepackage{fancyhdr}       
\usepackage{graphicx}       
\usepackage{amsmath}
\usepackage{amsthm}
\usepackage{bm}
\usepackage{mathtools} 
\usepackage{float}

\usepackage{algorithm}
\usepackage{algpseudocode}
\usepackage{xcolor}
\usepackage{comment}

\graphicspath{{media/}}     

\theoremstyle{definition}

\newtheorem*{theorem*}{theorem}

\newtheorem*{definition*}{difinition}
\newtheorem{lemma}{lemma}
\newtheorem*{lemma*}{lemma}

\newtheorem*{prop*}{proposition}

\title{A Unified Unfolding and Slide-Vector Model for Asymmetric Dissimilarities via Clustered Lasso 
}

\author{
  Ryohei Moriya \\
  Graduate School of Culture and Information Science \\
  Doshisha University \\
  Kyoto\\
  \texttt{ryohei.moriya.ss@gmail.com} \\
   \And
  Kensuke Tanioka \\
  Department of Biomedical Sciences and Informatics \\
  Doshisha University \\
  Kyoto\\
  \texttt{ktanioka@mail.doshisha.ac.jp} \\
  \And
  Hiroshi Yadohisa \\
  Department of Culture and Information Science \\
  Doshisha University \\
  Kyoto\\
  \texttt{hyadohis@mail.doshisha.ac.jp} \\
}

\begin{document}
\maketitle

\begin{abstract}
Asymmetric multidimensional scaling (AMDS) methods represent asymmetric dissimilarity data in a low-dimensional space. Existing approaches include the highly flexible unfolding model and the parsimonious slide-vector model. However, a highly flexible model is not necessarily optimal for adequately representing asymmetric relationships, and a model with an intermediate level of complexity may be more effective. In this study, we propose a new AMDS model that integrates the unfolding and slide-vector models by incorporating a clustered Lasso penalty. By fusing similar slide vectors, the proposed method can represent intermediate model structures and adjust model complexity according to the observed data. In numerical simulations, we compared the proposed method with existing methods by varying the number of groups, the number of objects, and the number of dimensions. The results demonstrated that the proposed method achieved high estimation accuracy across various asymmetric structures. Furthermore, in an application to cola brand-switching data, the proposed method achieved the highest accuracy in reproducing asymmetric relationships among the three methods compared and yielded a lower stress value than the slide-vector model. These findings suggest that adaptively adjusting model complexity may be effective for representing asymmetric relationships more accurately.
\end{abstract}


\section{Introduction}
Relationships among objects are often represented by dissimilarity data, which quantify the degree of dissimilarity between pairs of objects. Analyzing dissimilarity data enables researchers to reveal the underlying structures and relationships among objects, and such data have therefore been widely used in various fields, including psychology and marketing. For example, in psychology, perceptual structures have been analyzed based on similarity judgments of visual objects \cite{hout2014mm}. In marketing, brand similarity assessments have been used to analyze market structures, while brand-switching data have been used to investigate consumer behavior \cite{desarbo1992new}.

A representative method for analyzing dissimilarity data is Multidimensional Scaling (MDS) \cite{shepard1962analysis,kruskal1964nonmetric}. MDS is a technique for visualizing relationships among objects by mapping them into a low-dimensional space while preserving the observed dissimilarities as faithfully as possible. In the resulting configuration, the distances between objects represent their dissimilarities, thereby facilitating the interpretation of similarities and underlying structures. Although MDS is a classical method, it remains an active research topic, with recent studies focusing on large-scale data analysis and computational efficiency \cite{delicado2025multidimensional}, neural network-based approaches \cite{canzar2024metric}, and consumer decision-aware multidimensional scaling \cite{bijmolt2021adaptive}.

In practice, however, dissimilarity data are not always symmetric. Asymmetric relationships arise in a variety of fields, and various methods have been developed for analyzing such asymmetric structures \cite{saito2004data,okada2024applied}. For example, in psychology, similarity judgments based on human cognition are often asymmetric \cite{tversky1977features}. In marketing, brand-switching data provide another representative example, where the number of consumers switching from Brand A to Brand B generally differs from that in the opposite direction \cite{desarbo1992new,okada2012asymmetric}. Such asymmetric relationships cannot be adequately represented by conventional MDS, which assumes symmetric dissimilarities.

To address this limitation, numerous asymmetric multidimensional scaling (AMDS) methods have been proposed for directly analyzing asymmetric dissimilarity data \cite{borg2005modern,mair2022more,chino2025structure,saito2004data,okada2024applied}. AMDS has also been applied to practical problems involving brand-switching data, including the analysis of switching among existing brands and the evaluation of newly introduced brands \cite{okada2012asymmetric,okada2014evaluating}. One representative approach is the unfolding model \cite{coombs1964theory}, which provides considerable flexibility by representing each object with two points in the configuration space. However, this flexibility comes at the cost of a large number of parameters, resulting in a relatively complex model. To simplify the unfolding model, Zielman and Heiser proposed the slide-vector model \cite{zielman1993analysis}, which represents asymmetry using a single slide vector shared by all objects. Although this model is more parsimonious and easier to interpret, its assumption of a common slide vector may be too restrictive to capture diverse asymmetric relationships that require object-specific slide vectors.

Thus, these AMDS models exhibit a trade-off between model flexibility and complexity. However, a highly flexible model is not necessarily optimal for adequately representing asymmetric relationships, and a model with an intermediate level of complexity between the two may be more effective. Therefore, in this study, we propose a new extended model that incorporates a clustered Lasso \cite{she2010sparse} to combine the unfolding and slide-vector models, allowing the model complexity to be adjusted based on the observed data.

\section{Proposed method}
This section describes the proposed method. Specifically, we first explain the model and present the optimization problem. therefore, we  describe the algorithm for the proposed method.

\subsection{Proposed model}
In this subsection, we present an alternative formulation of the unfolding and introduce the proposed model.

Let   $\bm{\Delta}=(\delta_{ik})\in\mathbb{R}^{n\times n}$ is the asymmetric dissimilarity data, where $\delta_{ik}$ denotes the dissimilarity from object $i$ to object $k$. We assume that $\delta_{ik}\neq\delta_{ki}$ in general. Let $\bm{X}=(\bm{x}_1,\bm{x}_2,\ldots,\bm{x}_n)^{\top}$ and $\bm{Y}=(\bm{y}_1,\bm{y}_2,\ldots,\bm{y}_n)^{\top}$ denote two coordinate matrices, where $\bm{x}_i,\bm{y}_i\in\mathbb{R}^{d}$ $(i=1, 2, \ldots,n)$.

To propose an extended model that includes both the unfolding\cite{coombs1964theory} and the slide vector model\cite{zielman1993analysis}, we introduce multiple slide vectors into the unfolding .
We propose the following model:
\begin{equation}\label{equ:unfolding another expression}
    \delta_{ik}=\|\bm{x}_i-(\bm{x}_k-\bm{z}_k)\|_2+e_{ik}\quad (i,k=1,2,\cdots,n)
\end{equation}
where $e_{ik}\in\mathbb{R}$ denotes the error term, $d$ is the embedding dimension, and $\|\cdot\|_2$ denotes the Euclidean norm.

For arbitrary $\bm{x}_k,\bm{y}_k\in\mathbb{R}^{d}$, there exists $\bm{z}_k\in\mathbb{R}^{d}$ such that $\bm{y}_k=\bm{x}_k-\bm{z}_k$. Therefore, Equation~(\ref{equ:unfolding another expression}) is equivalent to the unfolding model. Furthermore, if $\bm{z}_1=\bm{z}_2=\cdots=\bm{z}_n$, Equation~(\ref{equ:unfolding another expression}) reduces to the slide-vector model.

\subsection{Optimization problem}
In this subsection, we show the optimization problem for an asymmetric multidimensional scaling method corresponding to the unfolding method and an extension of the slide-vector model. 

The optimization problem (\ref{equ:clusterd lasso unfolding}) is defined based on model (\ref{equ:unfolding another expression}).

\begin{equation}\label{equ:clusterd lasso unfolding}
    L(\bm{X}, \bm{Z}) = \sum_{i\neq k} (\delta_{ik} - \omega_{ik} \|\bm{x}_i - (\bm{x}_k - \bm{z}_k)\|_2)^{2} + \lambda \sum_{i<k} \|\bm{z}_i - \bm{z}_k\|_2
\end{equation}
Here, $\bm{Z}=(\bm{z}_1,\bm{z}_2,\cdots,\bm{z}_n)^{\top}$, $\bm{W}=(\omega_{ik})\quad(i,k=1,2,\cdots,n)$ is the weight matrix, and $\lambda(\geq0)$ is the tuning parameter. Here, the second term is the clustered Lasso term, which serves as a constraint to bring the slide vectors of each object closer together. When $\lambda=0$, equation (\ref{equ:clusterd lasso unfolding}) is equivalent to the optimization problem of the unfolding method, and when $\lambda$ is sufficiently large, equation (\ref{equ:clusterd lasso unfolding}) is equivalent to the optimization problem of the slide vector model.

Next, we reformulate Equation (\ref{equ:clusterd lasso unfolding}) in matrix form as shown in Equation (\ref{equ:clusterd lasso unfolding matrix}).
\begin{equation}\label{equ:clusterd lasso unfolding matrix}
    L(\bm{B})=\frac{1}{2}\|{\bm{W}^{\dagger}\odot\{}\bm{\Delta}^{\dagger}-\bm{D}(\bm{GB}){\}}\|_{F}^{2}+\lambda\sum_{i<k}\|\bm{h}_{ik}^{\top}\bm{B}\|_2
\end{equation}
where
\begin{equation}
\boldsymbol{\Delta}^{\dagger}=\left[\begin{array}{cc}
\boldsymbol{O}_{n \times n} & \boldsymbol{\Delta} \\
\boldsymbol{\Delta}^{\top} & \boldsymbol{O}_{n \times n}
\end{array}\right], \quad \boldsymbol{W}^{\dagger}=\left[\begin{array}{cc}
\boldsymbol{O}_{n \times n} & \boldsymbol{W} \\
\boldsymbol{W}^{\top} & \boldsymbol{O}_{n \times n}
\end{array}\right], \quad \boldsymbol{G}=\left[\begin{array}{cc}
\boldsymbol{I}_n & \boldsymbol{O}_{n \times n} \\
\boldsymbol{I}_n & -\boldsymbol{I}_n
\end{array}\right] \quad \boldsymbol{B}=\left[\begin{array}{c}
\boldsymbol{X} \\
\boldsymbol{Z}
\end{array}\right]
\end{equation}
where $\bm{O}_{n\times n}$ is an $n \times n$ square matrix with all elements equal to $0$, $\bm{I}_n$ is the $n \times n$ identity matrix, and $\bm{D}(\bm{GB})$ is the
matrix $\bm{GB}$ is the distance matrix. Furthermore, let $\bm{h}_{ik}$ be a vector of length $2n$ such that the $(n + i)$th element is $1$, the $(n + k)$th element is $-1$, and all other elements are $0$.
By Equation (\ref{equ:clusterd lasso unfolding matrix}), the proposed method can be formulated as an optimization problem for regularized symmetric multidimensional scaling with missing values ,
and can thus be formulated as a problem of estimating $\bm{B}$.

\subsection{Algorithm}
This section describes the optimization algorithm and update equations of the proposed method. The proposed method is optimized using a SMACOF-based majorization algorithm\cite{borg2005modern}, which is an instance of the MM (Majorization–Minimization) algorithm\cite{touw2022convex}. First, the majorizing function of the proposed objective function is given by

\begin{lemma} Majorizing function of the proposed method\\
Given $\bm{\Delta}^{\dagger}$ and $\bm{B}^{(t)}$, the majorizing function $M(\bm{B}|\bm{B^{(t)}})$ for any $\bm{B}$ is given by:
    \begin{equation}\label{equ:majorizing func}
        M(\bm{B}|\bm{B^{(t)}})=\frac{1}{2}\mathrm{tr}(\bm{B}^{\top}\bm{G}^{\top}\bm{V}\bm{GB}) -\mathrm{tr}\left\{\bm{B}^{\top}\bm{G}^{\top}\bm{C}(\bm{GB}^{(t)})\bm{GB}^{(t)}\right\} +\lambda\mathrm{tr}\left\{\bm{B}^{\top}\left(\sum_{i<k}\frac{1}{2\|\bm{h}_{ik}^{\top}\bm{B}^{(t)}\|}\bm{h}_{ik}\bm{h}_{ik}^{\top}\right)\bm{B}\right\}
    \end{equation}
    where $\bm{V}=\sum_{i<j}\omega_{ij}^{\dagger}\bm{A}^{(ij)},\quad \boldsymbol{C}\left(\boldsymbol{G} \boldsymbol{B}^{(t)}\right)=\left(c_{i j}\right) $
    \begin{align*}
& c_{i j}=\left\{\begin{array}{cc}
-\frac{w_{i j} \delta_{i j}}{d_{i j}(\mathbf{G B})} & \text { for } i \neq j \text { and } d_{i j}(\mathbf{G B}) \neq 0 \\
0 & \text{for } i \neq j \text{ and } d_{i j}(\mathbf{G B})=0
\end{array}\right. \\
& c_{i i}=-\sum_{j=1, j \neq i}^n c_{i j}
\end{align*}
Where, $A = (a_{ij})\in \mathbb{R}^{(n\times n)}$, $A^{(ij)}$ is a matrix with $a_{ii} = a_{jj} = 1, a_{ij} = a_{ji} = -1$, all other elements equal to $0$, and $\bm{W}^{\dagger}=(\omega_{ik}^{\dagger})\,(i,k=1,2,\cdots,2n),\bm{D}(\bm{GB})=(d_{ij}(\bm{GB}))\quad (i,j=1,2,\cdots,2n)$.
\end{lemma}
\begin{proof}
Applying the Cauchy–Schwarz inequality to the first term of Equation (\ref{equ:clusterd lasso unfolding matrix}) yields Equation (\ref{equ:cathy}).
\begin{equation}\label{equ:cathy}
    L(\bm{B}) \leq \frac{1}{2}\mathrm{tr}(\bm{B}^{\top}\bm{G}^{\top}\bm{V}\bm{GB})-\mathrm{tr} {\{}\bm{B}^{\top}\bm{G}^{\top}\bm{C}(\bm{G} {\bm{B}}^{(t)})\bm {G} {\bm{B} ^{(t)}\}}+\lambda\sum_{i<k}\|\bm{h}_{ik}^{\top}\bm{B}\|_2
\end{equation}
Next, applying the arithmetic-geometric mean formula to the third term of equation (\ref{equ:cathy}) yields equation (\ref{equ:AMGM}).
\begin{align}
    &\leq\frac{1}{2}\mathrm{tr}(\bm{B}^{\top}\bm{G}^{\top}\bm{V}\bm{GB})-\mathrm{tr} {\{}\bm{B}^{\top}\bm{G}^{\top}\bm{C}(\bm{G} {\bm{B}}^{(t)})\bm {G} {\bm{B} ^{(t)}\}} +\ \lambda\sum_{i<k}\left( \frac{1}{2\|\bm{h}_{ik}^{\top}\bm{B}^{(t)}\|}\|\bm{h}_{ik}^{\top}\bm{B}\|^2\right)\nonumber\\ \label{equ:AMGM}
    &=
\frac{1}{2}\mathrm{tr}(\bm{B}^{\top}\bm{G}^{\top}\bm{V}\bm{GB})-\mathrm{tr} {\{}\bm{B}^{\top}\bm{G}^{\top}\bm{C}(\bm{G} {\bm{B}}^{(t)})\bm {G} {\bm{B} ^{(t)}\}}+\lambda\mathrm{tr}\left\{\bm{B}^{\top}\left(\sum_{i<k}\frac{1}{2\|\bm{h}_{ik}^{\top}\bm{B}^{(t)}\|}\bm{h}_{ik}\bm{h}_{ik}^{\top}\right)\bm{B}\right\}
\end{align}
Here, the equality holds when $\bm{B}=\bm{B}^{(t)}$.
\end{proof}

The update equation for the proposed method can be expressed as follows using the majorizing function (\ref{equ:majorizing func}).

\begin{lemma}Update equation for $\bm{B}$\\
Given $\bm{V}$ and $\bm{B}^{(t)}$, the update equation for $\bm{B}$ is expressed as follows.
\begin{equation}\label {iter equ}
    \bm{B}=\left\{\bm{G}^{\top}\bm{VG}+\lambda\left(\sum_{i<k}\frac{1}{\|\bm{h}_{ik}^{\top}\bm{B}^{(t)}\|}\bm{h}_{ik}\bm{h}_{ik}^{\top}\ \right)\right\}^{-1}\bm{G}^{\top}\bm{C(\bm{GB}^{(t)})\bm{GB}^{(t)}}
\end{equation}
\end{lemma}
\begin{proof}
Since the majorizing function (\ref{equ:majorizing func}) is expressed entirely in terms of quadratic forms, the optimal solution occurs when the derivative of the majorizing function (\ref{equ:majorizing func}) is $0$. Therefore, the update equation for the proposed method is expressed as follows.
\begin{align}
M(\bm{B}|\bm{B}^{(t)})&=\bm{G}^{\top}\bm{VGB}-\bm{G}^{\top}\bm{C(\bm{GB}^{(t)})\bm{GB}^{(t)}} +\lambda\left(\sum_{i<k}\frac{1}{\|\bm{h}_{ik}^{\top}\bm{B}^{(t)}\|}\bm{h}_{ik}\bm{h}_{ik}^{\top}\right)\bm{B}=\bm{0}\\
\Longleftrightarrow 
    \bm{B}&=\left\{\bm{G}^{\top}\bm{VG}+\lambda\left(\sum_{i<k}\frac{1}{\|\bm{h}_{ik}^{\top}\bm{B}^{(t)}\|}\bm{h}_{ik}\bm{h}_{ik}^{\top}\ \right)\right\}^{-1}\bm{G}^{\top}\bm{C(\bm{GB}^{(t)})\bm{GB}^{(t)}}
\end{align}
\end{proof}

 {
Then, repeat the update using the update formula \ref{iter equ} until the stopping condition \ref{stop condition} is satisfied.
\begin{equation}\label{stop condition}
    \dfrac{
    L(\bm{B}_m^{(t-1)})-L(\bm{B}_m^{(t)})
    }{
    L(\bm{B}_m^{(t-1)})
    }<\varepsilon
\end{equation}
where $\varepsilon\,(>0)$ is tolerance .
}
 {
The proposed objective function is nonconvex with respect to the configuration parameters. Consequently, the MM algorithm is not guaranteed to attain a global minimizer, and the resulting solution may depend on the initial configuration. Such initialization dependence and the occurrence of local minima are well known in MDS \cite{borg2017choice}. To mitigate this initialization dependence, we employ a multi-start strategy.}

 {
The proposed method is run from $m$ randomly generated initial configurations, and the solution attaining the smallest value of the penalized objective function is retained.
}

 {

The MM algorithm is then applied to each of the $M$ initial configurations, and among the resulting $M$ solutions$\bm{\hat{B}}_{m}\quad (m=1,2,\dots,M)$, the solution that minimizes the penalized objective function is selected as the final estimate$\hat{\bm{B}}$.}

Algorithm 1 illustrates the proposed MM algorithm.
\begin{algorithm}[H]
\caption{Proposed method with multiple starts}
\label{alg:multistart}
\begin{algorithmic}[1]
\State \textbf{Input:} Dissimilarity matrix $\bm{\Delta}$,
tuning parameter $\lambda$, number of starts $M$ and
tolerance $\varepsilon$
\State \textbf{Output:} $\hat{\bm{B}}$

\For{$m = 1,2,\cdots,M$}
    \State Generate an initial configuration $\bm{B}_m^{(0)}$
    \State Set $t \gets 0$

    \Repeat
        \State Compute $\bm{C}(\bm{G}\bm{B}_m^{(t)})$
        \State Update $\bm{B}_m^{(t+1)}$ using Eq.~(\ref{iter equ})
        \State $t \gets t+1$
    \Until the stopping conditions(\ref{stop condition}) are satisfied

    \State $\hat{\bm{B}}_m \gets \bm{B}_m^{(t)}$
\EndFor

\State $m^* \gets
\underset{m=1,2, \cdots,M}{\arg\min}
L(\hat{\bm{B}}_m)$
\State $\hat{\bm{B}} \gets \hat{\bm{B}}_{m^*}$

\end{algorithmic}
\end{algorithm}

\subsection{Selecting Tuning Parameters}
Next, the tuning parameters for the clustered Lasso term included in the proposed method were
selected using a cross-validation method that intentionally masks a portion of the asymmetric dissimilarity matrix.
 {
In this study,
 The tuning parameter is then selected from these candidate values by cross-validation.
}
Next, $\alpha$\% of all elements in the asymmetric dissimilarity matrix were randomly masked, and estimation was performed. When $(i,k)$ was masked, $(k,i)$ was also masked.
 { Using the asymmetric dissimilarities reconstructed from the estimated configuration and the true asymmetric dissimilarity data,
we evaluated the reproducibility of the original asymmetric relationships in the masked regions.
As an evaluation metric for cross-validation,
we used Equation (\ref{evaluate_cv}).
\begin{equation}\label{evaluate_cv}
A_{CV}=
\sum_{(i,k)\in C}\left\{\bm{1}_A\left(\delta_{i k}>\delta_{k i}\right) \bm{1}_A\left(\Tilde{d}_{i k}(\bm{X},\bm{Z})>\Tilde{d}_{k i}(\bm{X},\bm{Z})\right)\right\}
\end{equation}
where
$\Tilde{d}_{ik}(\bm{X},\bm{Z})
=\|\bm{x}_i-(\bm{x}_k-\bm{z}_k)\|_2$,
$\bm{1}_{A}(\cdot)$ denotes the indicator function,
and the subset of pairs of masked objects is denoted by
$C\subset\{1,2,\ldots,n\}\times\{1,2,\ldots,n\}$.
A larger value indicates higher accuracy. Furthermore, to reduce the influence of the masking position,}
we conducted $R$ validation runs for each tuning parameter by randomly varying the masking position,
and used the average$\bar{A}_{CV}$ of these results as the evaluation value for that tuning parameter.

Algorithm 2 illustrates the selection of the tuning parameter.
\begin{algorithm}[t]
\caption{Selection of the tuning parameter by cross-validation}
\label{alg:cv}
\begin{algorithmic}[1]

\Statex \textbf{Input:} Dissimilarity matrix $\bm{\Delta}$,
candidate tuning parameters
$\{\lambda_1,\ldots,\lambda_Q\}$,
number of CV repetitions $R$,
hold-out ratio $\eta$, and number of starts $M$

\Statex \textbf{Output:} Selected tuning parameter $\hat{\lambda}$

\For{$r=1,2,\ldots,R$}
    \State Randomly select a set of unordered held-out pairs
    $C_r$ with hold-out ratio $\eta$
    \State Construct the weight matrix $\bm{W}_r$
    by masking both $(i,k)$ and $(k,i)$
    for each $(i,k)\in C_r$
\EndFor

\For{$q=1,2,\ldots,Q$}

    \For{$r=1,2,\ldots,R$}

        \State Estimate $\hat{\bm{B}}_{r,q}$
        using Algorithm~\ref{alg:multistart}
        with $\lambda_q$ and $\bm{W}_r$

        \State Compute the estimated dissimilarities
        $\tilde{d}_{ik}(\hat{\bm{X}}_{r,q},
        \hat{\bm{Z}}_{r,q})$
        and
        $\tilde{d}_{ki}(\hat{\bm{X}}_{r,q},
        \hat{\bm{Z}}_{r,q})$
        for $(i,k)\in C_r$

        \State Compute
        $A_{\mathrm{CV}}^{(r)}(\lambda_q)$
        using Eq.~(\ref{evaluate_cv})

    \EndFor

    \State Compute the mean of the valid CV scores
    \[
    \bar{A}_{\mathrm{CV}}(\lambda_q)
    =
    \frac{1}{R}
    \sum_{r=1}^{R}
    A_{\mathrm{CV}}^{(r)}(\lambda_q)
    \]

\EndFor

\State Select
\[
\hat{\lambda}
=
\underset{\lambda}{\arg\max}\left\{
\bar{A}_{\mathrm{CV}}(\lambda_q)
\right\}
\]

\end{algorithmic}
\end{algorithm}

\section{Numerical simulation}
For asymmetric data that share a common slide vector structure, we compare and evaluate the accuracy of the proposed method, unfolding , and the slide vector method in estimating asymmetric relationships.

\subsection{Simulation design}
In this simulation, the data is divided into multiple groups so that the number of data points in each group is as evenly distributed as possible, and data is generated such that data points belonging to the same group share a common slide vector. By varying the number of groups, we evaluate the performance of unfolding, the slide-vector method, and the proposed method with respect to the data structures assumed by each approach. Furthermore, by varying the sample size and the number of dimensions, we compare the estimation accuracy of asymmetric relationships and the overall model fit for each method under these different conditions.

The simulation data were generated according to the following procedure:

\begin{enumerate}
    \item Generate the object configuration matrix $\bm{X}\in\mathbb{R}^{n\times d}$ from the multivariate normal distribution $\mathcal{N}(\bm{0},\bm{I})$, where $\bm{I}$ denotes the identity matrix of order $d$.

    \item Divide the objects into clusters. First, define the membership matrix
    $\bm{G}=(g_{i\ell})\in \{0,1\}^{n\times L}$, where $\sum_{l=1}^{L}g_{i\ell}=1\quad (i=1,2,\cdots,n)$, and $L$ represents the number of clusters. Here, $g_{i\ell}=1$ indicates that object $i$ belongs to cluster $\ell$.

    \item Next, let $\bm{S}=(\bm{s}_{1},\bm{s}_{2},\cdots,\bm{s}_L)^{\top}\in \mathbb{R}^{L\times d}$ be the slide vectors for each cluster. The slide vectors for each cluster are generated from a multivariate normal distribution $\mathcal{N}(\bm{0}, 0.3^2\bm{I})$. Finally, we generate the slide vectors $\bm{Z}=\bm{GS}\in \mathbb{R}^{n\times d}$ for each object.

    \item Generate the asymmetric dissimilarity matrix $\bm{\Delta}$ by computing
    $\delta_{ik}=\|\bm{x}_i-(\bm{x}_k-\bm{z}_k)\|+e_{ik}$
    for all pairs of objects $(i,k)$, where $e_{ik}$ are independently generated from the normal distribution $\mathcal{N}(0,\sigma^2)$ with $\sigma=0.2$.
\end{enumerate}
In this study,
we will conduct verification on a total of 30 cases where $n \in \{20, 50, 100\}$, $d \in \{2, 3\}$, and $L \in \{1, 2, 3, 5, n\}$. A total of 100 trials were performed for each case.For cross-validation, we set $\alpha = 30,M=10$ and $R = 5$.

For the final model evaluation,
we used Equation (\ref{evaluate}).
\begin{equation}\label{evaluate}
A_{\mathrm{dir}}=\frac{\sum_{(i,k)\in C_0}\left\{\bm{1}_A\left(\delta_{i k}>\delta_{k i}\right) \bm{1}_A\left(\Tilde{d}_{i k}(\bm{X},\bm{Z})>\Tilde{d}_{k i}(\bm{X},\bm{Z})\right)\right\}}{(n^2-n)/ 2}
\end{equation}
Where $C_0 = \{1, 2, \cdots, n\} \times \{1, 2, \cdots, n\}$. A value greater than 0, the closer to 1, indicates that the asymmetric relationship is correctly reproduced.

Furthermore, the overall fit of the model is evaluated using the stress measure $\sum_{i\neq k}\|\delta_{ik}-\hat{d}_{ik}\|$. This measure takes values of 0 or greater, and the closer it is to 0, the better the model fits the input asymmetric dissimilarity data.

\subsection{ {Simulation result}}
\begin{figure}[H]
\begin{center}
\includegraphics[scale=0.46]{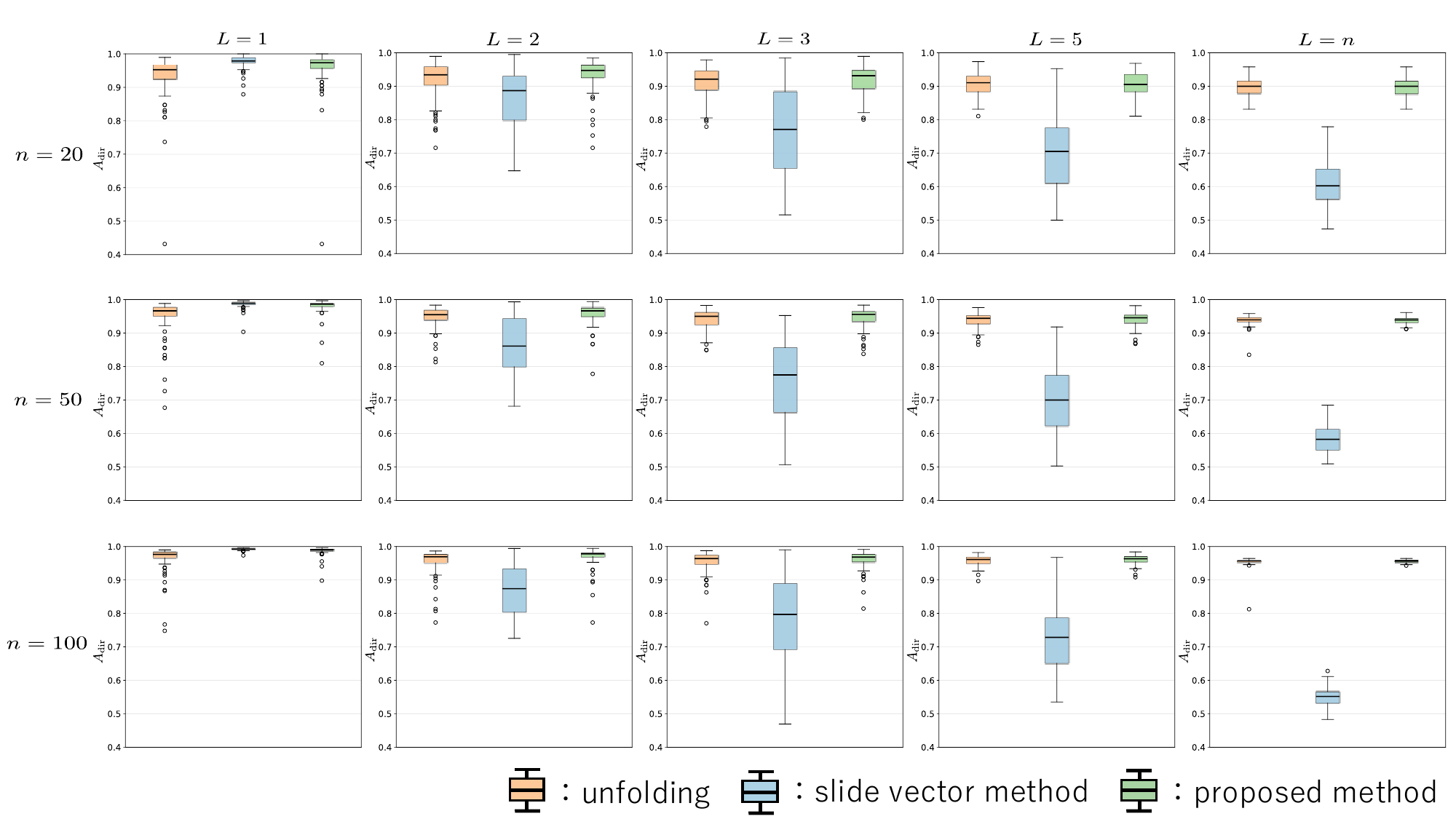}
\caption{Boxplots of $A_{\mathrm{dir}}$ for each method under different simulation settings with $d=2$. The rows correspond to the sample sizes $n$, and the columns correspond to the numbers of clusters $L$. The vertical axis represents $A_{\mathrm{dir}}$, and the horizontal axis represents the methods.}
\label{assinmetric acuracy boxplot d2} 
\end{center}
\end{figure}

\begin{figure}[H]
\begin{center}
\includegraphics[scale=0.46]{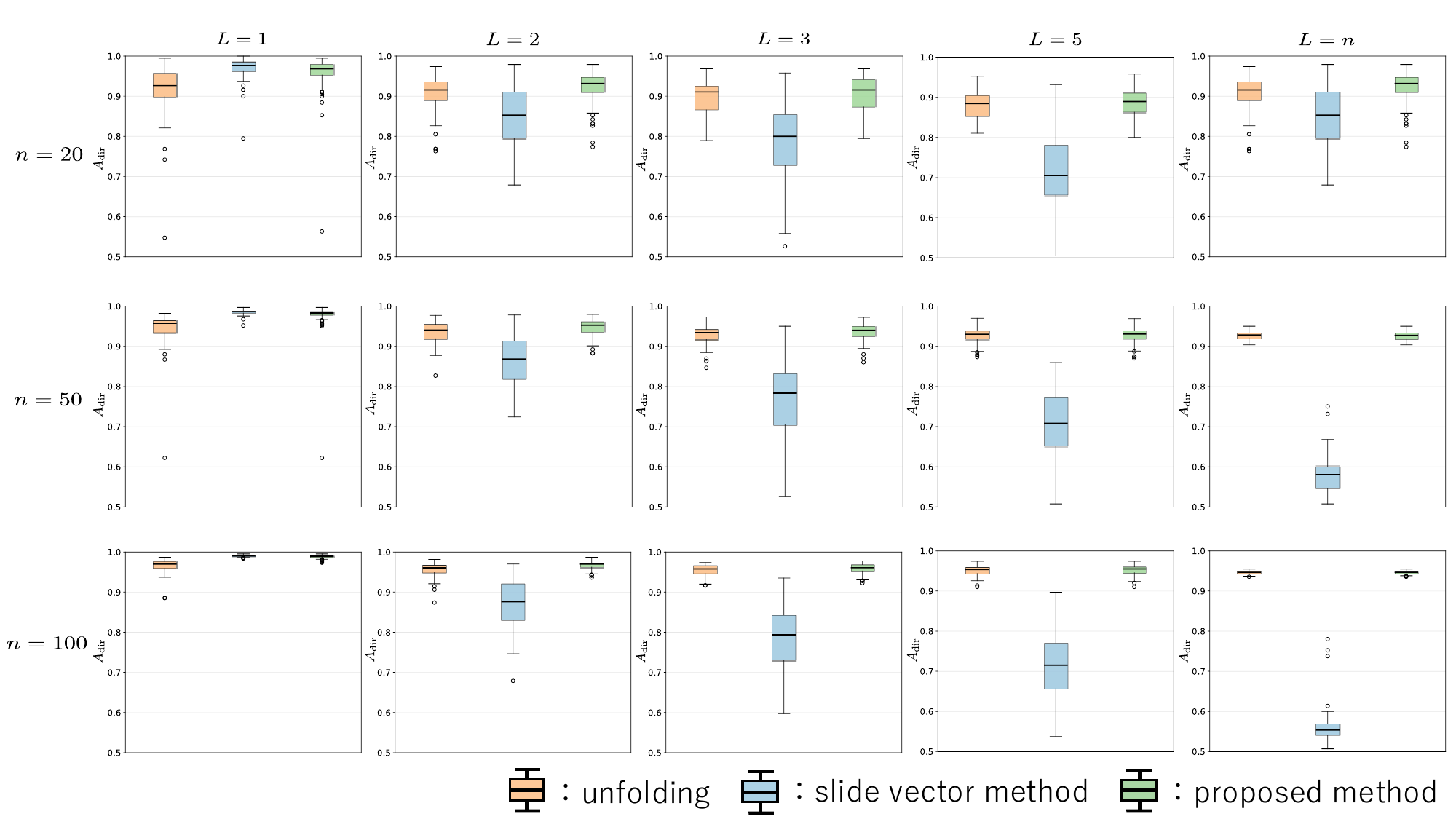}
\caption{Boxplots of $A_{\mathrm{dir}}$ for each method under different simulation settings with $d=3$. The rows correspond to the sample sizes $n$, and the columns correspond to the numbers of clusters $L$. The vertical axis represents $A_{\mathrm{dir}}$, and the horizontal axis represents the methods.}
\label{assinmetric acuracy boxplot d3} 
\end{center}
\end{figure}

\begin{figure}[H]
\begin{center}
\includegraphics[scale=0.46]{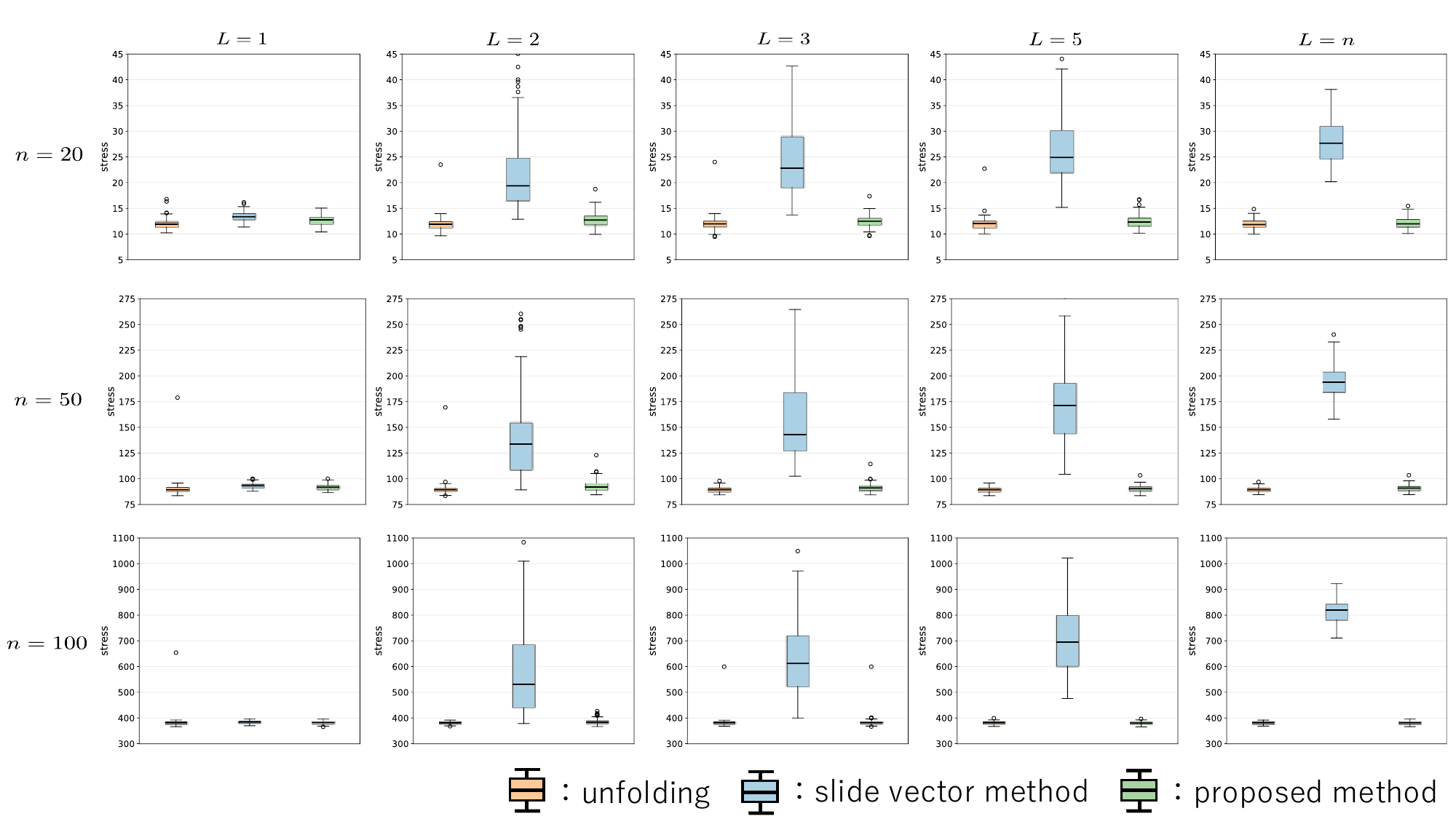}
\caption{Boxplots of $stress$ for each method under different simulation settings with $d=2$. The rows correspond to the sample sizes $n$, and the columns correspond to the numbers of clusters $L$. The vertical axis represents $stress$, and the horizontal axis represents the methods.}
\label{stress boxplot d2} 
\end{center}
\end{figure}

\begin{figure}[H]
\begin{center}
\includegraphics[scale=0.46]{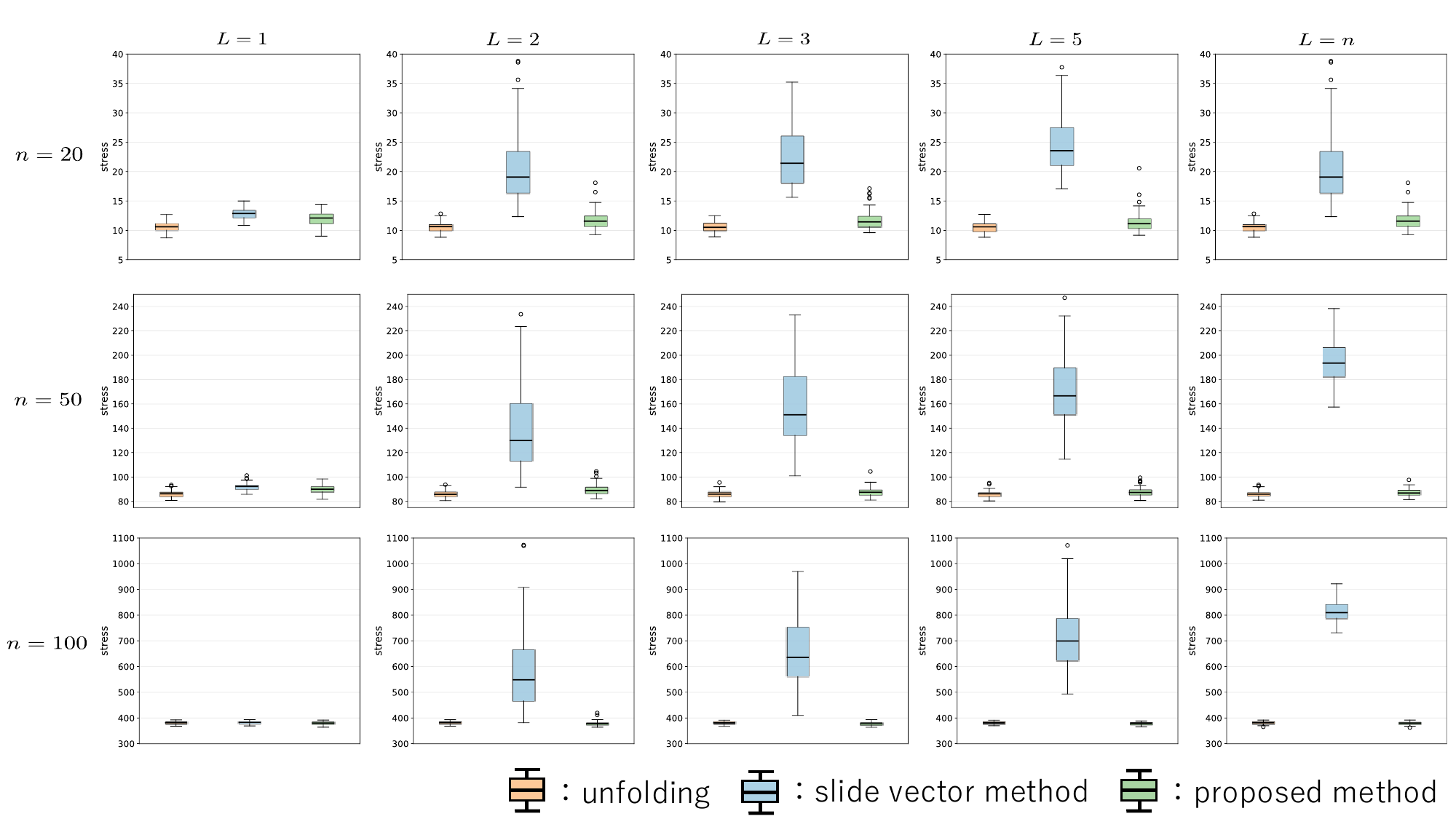}
\caption{Boxplots of $stress$ for each method under different simulation settings with $d=3$. The rows correspond to the sample sizes $n$, and the columns correspond to the numbers of clusters $L$. The vertical axis represents $stress$, and the horizontal axis represents the methods.}
\label{stress boxplot d3} 
\end{center}
\end{figure}

This section presents the results of the numerical simulations. Figs.~\ref{assinmetric acuracy boxplot d2} and \ref{assinmetric acuracy boxplot d3} show the accuracy $A_{\mathrm{dir}}$ in reproducing the directions of asymmetric relationships obtained by the unfolding model, the slide-vector model, and the proposed method for $d=2$ and $d=3$, respectively. Figs.~\ref{stress boxplot d2} and \ref{stress boxplot d3} show the corresponding stress values for $d=2$ and $d=3$, respectively.

First, we examine the accuracy in reproducing the directions of asymmetric relationships. Focusing on the median values of $A_{\mathrm{dir}}$ in Figs.~\ref{assinmetric acuracy boxplot d2} and \ref{assinmetric acuracy boxplot d3}, the proposed method generally achieved high accuracy across different numbers of objects and groups. In particular, for the intermediate group structures with $\ell=2$, $3$, and $5$, the proposed method achieved the highest median accuracy among the three methods under most conditions. However, when $n=20$, $d=2$, and $\ell=5$, the unfolding model achieved a higher median accuracy than the proposed method.

When $\ell=1$, which corresponds to the structure assumed by the slide-vector model, the slide-vector model achieved the highest accuracy, followed by the proposed method. In contrast, when $\ell=n$, which corresponds to the structure assumed by the unfolding model, the unfolding model and the proposed method achieved comparable accuracy, both outperforming the slide-vector model. These tendencies were observed for both $d=2$ and $d=3$.

These results indicate that the proposed method can flexibly accommodate various data structures, ranging from a single common slide vector shared by all objects to distinct slide vectors for individual objects. In particular, although the proposed method did not outperform the slide-vector model when $\ell=1$, it maintained high accuracy under this setting. Moreover, when $\ell=n$, the proposed method achieved accuracy comparable to that of the unfolding model. These findings suggest that the proposed method provides a flexible framework capable of accommodating the structural assumptions of both conventional models while also representing intermediate group structures through the fusion penalty.

Next, we examine the overall model fit using the stress values shown in Figs.~\ref{stress boxplot d2} and \ref{stress boxplot d3}. Overall, the unfolding model tended to yield the lowest stress values, followed by the proposed method and the slide-vector model. However, when $n=100$, $d=3$, and $\ell=1$, the unfolding model yielded the lowest stress values, followed by the slide-vector model and the proposed method.

\section{ {Real example}}
In this section, we present the results of an analysis of brand-switching data for cola products as an example of applying the proposed method to real-world data. In this study, we apply an asymmetric multidimensional scaling method based on the clustered Lasso, select tuning parameters using cross-validation, and compare and evaluate the accuracy of reproducing the asymmetric structure of the intermediate model-selected via the unfolding method, the slide-vector model, and cross-validation.

n this study, we use the cola brand-switching data \cite{borg2005modern}, which were derived from supermarket scanner data \cite{bell1998shopping}. 
This dataset records the purchase history of 15 types of cola products
for 488 households in the United States over a period of 104 weeks,
from June 1991 to June 1993.
Here, a brand switch refers to
a situation where different products are selected in two consecutive purchases.
Brand switching data
represents the frequency of switching from one product to another;
since a higher value indicates that switching between two products is more likely to occur,
it can be interpreted as an indicator of competition between products. This switching data was converted into
asymmetric dissimilarity data based on a gravity model.

The model is evaluated using the accuracy of asymmetric relationship estimation, $A_{\mathrm{dir}}$ and the stress measure. In addition to we set $\alpha=30,R=20,M=100$

\begin{figure}[H]
\begin{center}
\includegraphics[width=\linewidth]{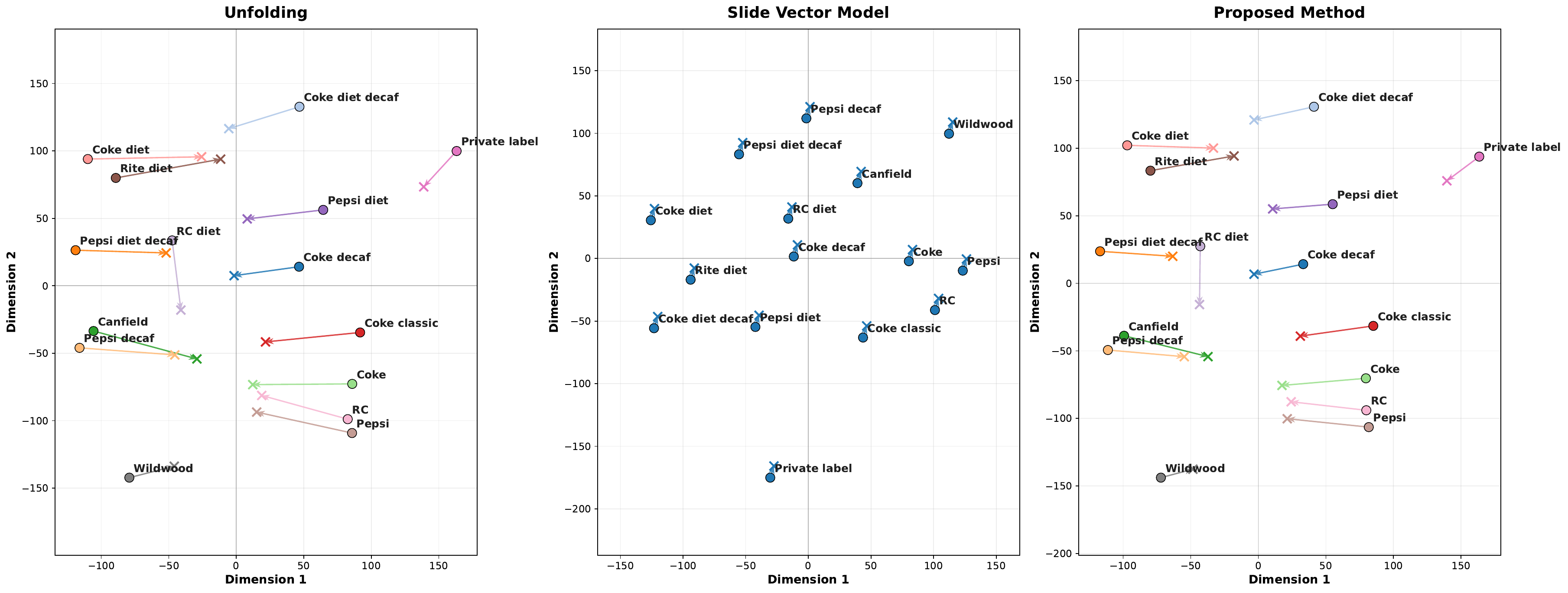}
\caption{Two-dimensional representations of the data using each method: The figure on the left shows the unfolding method, the middle figure shows the slide vector model, and the figure on the right shows the proposed method. Additionally, objects with similar slide vectors are grouped together and shown in the same color.}
\label{real data plot} 
\end{center}
\end{figure}

First, as shown in Table~\ref{tab:real_results}, the proposed method achieved the highest $A_{\mathrm{dir}}$, defined by Equation~(\ref{evaluate}), among the three methods. Regarding stress, the proposed method yielded a higher value than the unfolding model but a lower value than the slide-vector model. Figure~\ref{real data plot} shows the two-dimensional configurations obtained by the unfolding model, the slide-vector model, and the proposed method. The arrows represent the slide vectors, while the points represent the coordinates of the objects. Objects sharing the same slide vector are represented by points of the same color. No clear group structure was observed in the configuration obtained by the proposed method.
These results suggest that, in this real-data application, the proposed method reproduced the directions of asymmetric relationships more accurately than both the unfolding model and the slide-vector model, while maintaining a reasonable level of overall model fit, even though no clear group structure was identified.

Next, we examine the interpretation of the two-dimensional configuration obtained only by the proposed method. Table~\ref{tab:asym_relation} shows examples of asymmetric relationships that were correctly reproduced by the proposed method. In interpreting the configuration, the dissimilarity between two items reflects the magnitude of the outflow between them; thus, items located close to each other are considered to have a stronger competitive relationship. In addition, when an asymmetric relationship exists between two items, the item associated with the larger directional distance can be interpreted as having greater competitiveness.

Specifically, we focus on "Coke classic" and "Rite diet". The observed directional difference indicates an asymmetric
relationship from "Coke classic" to "Rite diet".The proposed method correctly reproduces this direction, whereas the unfolding method and the slide vector model estimate the opposite direction. Therefore, the proposed method allows us to identify the difference in competitiveness
between these two items, with "Coke classic" interpreted as being more competitive than "Rite diet".

We also consider "Coke diet" and "RC"
In this case, the observed directional difference indicates an asymmetric relationship from "RC" to "Coke diet".
Although the unfolding method and the slide vector model estimate the opposite direction, the proposed method correctly reproduces this relationship. This suggests that "RC" can
be interpreted as being more competitive than "Coke diet".

\begin{table}[H]
\centering
\caption{Comparison of estimation accuracy and stress for the real data.}
\label{tab:real_results}
\begin{tabular}{lccc}
\hline
 & Unfolding & Slide Vector Model & Proposed Method \\
\hline
$A_{\mathrm{dir}}$
& 0.583
& 0.573
& 0.602 \\
Stress
& 362495
& 432012
& 368737 \\
\hline
\end{tabular}
\end{table}

\begin{table}[H]
\centering
\caption{Regarding Asymmetric Relationships Correctly Estimated Only by the Proposed Method}
\label{tab:asym_relation}
\begin{tabular}{lll}
\hline
item i & item j &True direction \\
\hline
 Coke classic & Rite diet
& Coke classic $\leftarrow$Rite diet \\
Coke diet & RC
& Coke diet $\rightarrow$ RC\\
\hline
\end{tabular}
\end{table}

\section{ {Conclusion}}
In this study, we proposed a new asymmetric multidimensional scaling method incorporating a clustered Lasso penalty to balance model flexibility and interpretability in the analysis of asymmetric dissimilarity data. By fusing similar slide vectors, the proposed method represents a structure intermediate between the unfolding model and the slide-vector model, allowing model complexity to be adjusted according to the observed data.

Numerical simulations demonstrated that the proposed method accurately estimated asymmetric relationships across various group structures, ranging from structures in which all objects share a common slide vector to those in which each object has a distinct slide vector. Furthermore, in the application to brand-switching data, the proposed method achieved the highest $A_{\mathrm{dir}}$ among the methods compared and yielded a lower stress value than the slide-vector model. These results suggest that the proposed method can reproduce the directions of asymmetric relationships more accurately while maintaining a reasonable level of overall model fit.

One challenge for future research is improving computational efficiency. Since the proposed method may converge to different local solutions depending on the initial values, estimation is performed using multiple initial configurations. Furthermore, because cross-validation is used to select the tuning parameter, the computational cost increases as the number of objects increases. Future work should focus on reducing sensitivity to initial values and investigating more efficient optimization algorithms and tuning parameter selection methods.


\bibliographystyle{unsrt}  
\bibliography{references}

\end{document}